# Unified physical framework for stretched-exponential, compressed-exponential, and logarithmic relaxation phenomena in glassy polymers

Valeriy V. Ginzburg[1], Oleg V. Gendelman[2], and Alessio Zaccone[3]

[1]Department of Chemical Engineering and Materials Science, Michigan State University, East Lansing, Michigan, USA

[2]Faculty of Mechanical Engineering, Technion, Haifa, Israel

[3]Department of Physics, University of Milan, Milan, Italy

# Abstract

We develop a simple yet comprehensive nonlinear model to describe relaxation phenomena in amorphous glass-formers near the glass transition temperature. The model is based on the two-state, two-(time)scale (TS2) framework, and describes the isothermal relaxation of specific volume, enthalpy, or shear stress via a simple first-order nonlinear differential equation (the Trachenko-Zaccone [TZ] equation) for local cooperative events. These nonlinear dynamics of cooperatively rearranging regions (CRR) naturally arise from the TS2 framework. We demonstrate that the solutions of the TZ equation comprehensively encompass the Debye exponential relaxation, the Kohlrausch-Williams-Watts (KWW) stretched and compressed relaxations, and the Guiu-Pratt logarithmic relaxation. Furthermore, for the case of stress relaxation modeling, our model recovers, as one of its limits, the Eyring law for plastic flow, where the Eyring activation volume is related to thermodynamic parameters of the material. Using the example of polystyrene (PS), we demonstrate how our model successfully describes the Kovacs' "asymmetry of approach" specific volume and enthalpy experiments, as well as the stress relaxation. Other potential applications of the model, including the dielectric relaxation, are also discussed. The presented approach disentangles the physical origins of different relaxation laws within a single general framework based on the underlying physics.

## 1. Introduction

Relaxation (dielectric[1–5], mechanical[6–16], specific volume[17–25], or enthalpy[26–28]) experiments are important tools in understanding the behavior of glassy and near-glassy polymers.[3,29,30] Unlike conventional liquids, glassy materials exhibit relaxations that substantially deviate from the standard Debye single-exponential decay. In some cases, the material property of interest changes slowly over many decades in time ("stretched relaxation"), while in others, it exhibits no change until a sudden avalanche-like event takes place ("compressed relaxation"). Depending on the specific experiment, multiple fundamental and empirical theories and models have been proposed to describe those relaxation phenomena. Some of the more successful theories, such as the Kovacs-Aklonis-Hutchinson-Ramos[26] (KAHR) or the Tool-Narayanaswamy-Moynihan[31–33] (TNM) frameworks, linearize the relaxation process by introducing auxiliary variables like "material time" or "fictive temperature" – see, e.g., Dyre[34–36] et al. for a broader perspective on the concept of material time. On the other hand, from the practical standpoint, it was often useful to describe relaxations using simple analytical functions. In the context of dielectric spectroscopy, the Kohlrausch-Williams-Watts[37,38] (KWW) function, $y = \exp\left[-\left(t/\tau\right)^{\beta}\right]$, was proposed and shown to successfully capture many salient features of real-life relaxation spectra; furthermore, it was shown[39] that the Fourier transform of the KWW function corresponds to the well-known Havriliak-Negami[40,41] spectrum. The KWW function is characterized by the power exponent $\beta$ -- when 0<$\beta$<1, the relaxation is "stretched", when $\beta$=1, it is Debye (single-exponential), and when $\beta$>1, it is "compressed" (avalanche-like transition). At the same time, the KWW function is not smooth in the limit $t \rightarrow 0$, and its physical interpretation is often unclear. Another functional form (most often used in the context of stress relaxation) is the logarithmic dependence of the response variable on time (usually referred to as the Guiu-Pratt law[42,43]). The logarithmic dependence on the time was also observed empirically for enthalpy and volume relaxations (see, e.g., Kovacs[44] and Malek[45,46]) – this result can be reproduced in numerical solutions of the TNM or KAHR equations, but requires some fine-tuning of the model parameters.

Recently, Trachenko and Zaccone[47] (TZ) described a simple nonlinear relaxation equation, $\dot{y} = -\left(y/\tau\right)\exp(Ky)$, whose solutions can be approximated by the stretched or compressed KWW functions. Further analysis can show additional interesting features of the TZ equation. Unlike KWW, its solution is smooth at t → 0. Furthermore, at early times, its solution can be approximated by the logarithmic Guiu-Pratt function, while at late times, it approaches a single exponential. Finally, it is easy to re-cast in terms of material time – such an interpretation would be very similar to the recent analysis of Niss et al.[36]

If one is to use the TZ equation to describe the volume, enthalpy, or stress relaxation in polymeric glass-formers, the question arises – what is the physical meaning of the parameters used in this equation and how they can be tied to other material properties (preferably the equilibrium ones). Here, we demonstrate that the TZ relaxation equation can be derived in a straightforward way within the "two-state, two-(time)scale" (TS2) framework recently proposed by Ginzburg et al.[48–53] Within this framework, the "fast" state variable, $\psi$ (the "solid" state fraction), relaxes on the timescale of the beta-relaxation time, $\tau_\beta$. The "slow" state variable, $\nu$ (the lattice "occupancy"; [1-$\nu$] is the fractional free volume), relaxes on the timescale of the alpha-relaxation time, $\tau_\alpha$, which, in turn, depends on $\psi$. This approach builds on the ideas proposed by Ngai and co-workers (the "Coupling Model" [CM]),[54] as well as the recent work of Napolitano and co-workers (the "Slow Arrhenius Process" [SAP]).[55] On timescales significantly larger than $\tau_\beta$, we can then integrate out the fast variable $\psi$ and arrive at the TZ equation for the slow variable $\nu$. Furthermore, if we accept that the relaxation variables (stress, specific volume, or enthalpy) undergo only small perturbations, the relaxation equation for them is also the TZ equation. The two parameters of the TZ equation, $\tau = \tau_{\alpha, eq}$ (the equilibrium $\alpha$-relaxation time at the current temperature) and $K$ (the cooperativity parameter) can be derived in a straightforward way if the equation of state (EoS) for the material is known. Here, we illustrate our approach using the two-state Sanchez-Lacombe model, but other two-state EoS can be utilized as well.

## 2. The Model

### 2.1. Free Energy, Equation of State, and Dynamics

Our starting point is the two-state, two-(time)scale (TS2) lattice model.[48] Within this framework, the material is divided into cooperatively rearranging regions (CRR) with mass $M$ – it can be one or several repeat units (for polymers) or one or more molecules (for low-molecular weight glass-formers). Each CRR can be in a "Liquid" (L) or "Solid" (S) state, with the solid fraction defined as $\psi$. Additionally, some lattice sites could be unoccupied (voids or free volume), with the occupied fraction (by both solid and liquid states) defined as $v$. We assume that the free energy per CRR, $G$, of a glass-forming material at atmospheric pressure ($P \approx 0$) can be written as,

$$G = -\varepsilon^* \frac{r}{v} \left[ \left( \frac{v\psi r_S}{r} \right)^2 + 2\alpha_{LS} \left( \frac{v\psi r_S}{r} \right) \left( \frac{v\{1-\psi\}r_L}{r} \right) + \alpha_{LL} \left( \frac{v\{1-\psi\}r_L}{r} \right)^2 \right]$$

$$+ k_B T \left[ \psi \ln\left( \frac{v\psi r_S}{r} \right) + \{1-\psi\} \ln\left( \frac{v\{1-\psi\}r_L}{r} \right) + r\frac{1-v}{v} \ln(1-v) \right]$$

(1)

Here, $T$ is the absolute temperature, $k_B$ is the Boltzmann's constant, $r_S$ and $r_L$ are the number of lattice sites occupied by the "Solid" and "Liquid" states of the glass-former, respectively. The van-der-Waals interaction energies are $\varepsilon_{LL}$ ("Liquid"-"Liquid" nearest neighbors), $\varepsilon_{SS}$ ("Solid"-"Solid" nearest neighbors), and $\varepsilon_{LS}$ ("Liquid"-"Solid" nearest neighbors); we can then define $\alpha_{LL} = \varepsilon_{LL}/\varepsilon_{SS}$, $\alpha_{LS} = \varepsilon_{LS}/\varepsilon_{SS}$, and

$\varepsilon^* = Z\varepsilon_{SS}/2$ (where Z is the coordination number). Finally, $r = \psi r_S + [1-\psi]r_L$. For more details, see Ginzburg et al.[50,52]

Equation 1 is a special version of the generic TS2 approach where a Sanchez-Lacombe (SL) EoS is adopted.[50,52,56,57] Other possible EoS can be used, but two main assumptions for any TS2 model are as follows. First, as discussed above, the state of the material is described by two variables, $\psi$ -- the fraction of the slow, low-temperature ("Solid") state relative to the fast, high-temperature ("Liquid") state, and $\nu$ -- the "lattice occupancy" ([1- $\nu$] is the void or free volume fraction). Second, the characteristic relaxation time for $\psi$ is the "fast" beta-relaxation time $\tau_\beta$, while the characteristic relaxation time for $\nu$ is the "slow" alpha-relaxation time, $\tau_\alpha$. At some temperature $T = T_A$ ("Arrhenius temperature"), $\tau_\alpha(T)$ and $\tau_\beta(T)$ merge, and the temperature dependence of the relaxation time is described by the Arrhenius law (although reaching this region experimentally is usually very difficult as the materials tend to evaporate or degrade at those high temperatures). The dynamics of isothermal relaxation for $\psi$ and $\nu$ are thus given by,

$$\frac{d\psi}{dt} = \frac{\psi^*(\nu) - \psi}{\tau_\beta} \qquad (2a)$$

$$\frac{d\nu}{dt} = \frac{\nu^* - \nu}{\tau_\alpha(\psi)} \qquad (2b)$$

$$\tau_\beta = \tau_\infty \exp\left[\frac{E_1}{RT}\right] \qquad (2c)$$

$$\tau_\alpha = \tau_\infty \exp\left[\frac{E_1}{RT} + \frac{E_2 - E_1}{RT}\psi\right] \qquad (2d)$$

As before, $E_1$ and $E_2$ are the activation energies of the "liquid" and "solid" states, respectively, and $\tau_\infty$ is the "elementary time" of molecular processes. Also, we define $\psi^*(\nu)$ as the solution of equation

$\frac{\delta G(\psi,v)}{\delta \psi} = 0$, and $v^*$ as the solution of equation $\frac{\delta G(\psi^*(v),v)}{\delta v} = 0$. (Note that for a given temperature, T, $v^*$ is a single number, and $\psi^*(v)$ is a function.)

### 2.2. The Kovacs' "Up" and "Down" Temperature Jumps and the "Asymmetry of Approach" for the Specific Volume and Enthalpy Experiments

Here, we describe the application of the above framework to the problem of "asymmetry of approach".[17,18,22,24,26,28,29,58,59] In these experiments, the material is quenched from its initial temperature, $T_I$, to the final temperature, $T_a$, and is allowed to relax isothermally. The response measured is either the specific volume, $V(t)$, or the enthalpy, $H(t)$ (note that at $P = 0$, the enthalpy, $H$, is equal to the internal energy, $U$). Within the SL-TS2 framework, they are given by,

$$H \propto -\varepsilon^* \frac{r}{v}\left[\left(\frac{v\psi r_S}{r}\right)^2 + 2\alpha_{LS}\left(\frac{v\psi r_S}{r}\right)\left(\frac{v\{1-\psi\}r_L}{r}\right) + \alpha_{LL}\left(\frac{v\{1-\psi\}r_L}{r}\right)^2\right]$$

(3a)

$$V = v_{sp,0}\frac{r}{r_S v} = v_{sp,0}\frac{r_S\left[\psi + \left(\frac{r_L}{r_S}\right)(1-\psi)\right]}{r_S v}$$

$$= v_{sp,0}\frac{\left[1 + \left(\frac{r_L}{r_S}-1\right)(1-\psi)\right]}{v} \approx v_{sp,0}\frac{e^{q(1-\psi)}}{v}$$

(3b)

where $v_{sp,0}$ is the specific volume at T = 0 K, and $q = r_L/r_S - 1$.

Let us now assume that the rate of temperature change is sufficiently fast and therefore, the time when the temperature of the material reaches the final temperature $T_a$ (let us call it $\tau_{on}$) is on the order of or smaller than the alpha relaxation time at the initial temperature $T_I$, $\tau_\alpha(T_I)$, $\tau_{on} < \tau_\alpha(T_I)$. This means that the beta relaxation in our experiment is always "fast enough" so that we can integrate eq. 2a and assume that at any time, $\psi = \psi^*(v)$. Then, at the beginning of the experiment, at t = $\tau_{on} \approx 0$, $v = v^*(T_I) \equiv v_I$, and $\psi = \psi^*(v_I) \equiv \psi_I$ (note that $v^*(T_I) \equiv v_I$ corresponds to the equilibrium value of $v$ at $T = T_I$, while $\psi^*(v_I) \equiv \psi_I$ refers to partial equilibration at $T = T_a$). Substituting $\psi_I$ and $v_I$ into eqs. 3a and 3b, we can determine the initial values of H and V, labeled $H_I$ and $V_I$, respectively.

We define the "normalized specific volume deviation" (NSVD), $y_V$, and the "normalized enthalpy deviation" (NED), $y_H$, as follows,

$$y_V = \frac{V - V^*}{V_I - V^*} \tag{4a}$$

$$y_H = \frac{H - H^*}{H_I - H^*} \tag{4b}$$

Both NSVD and NED are functions of $v$, and we approximate them as linear functions,

$$y_V = \frac{dy_V}{dv}(v - v^*) \tag{5a}$$

$$y_H = \frac{dy_H}{dv}(v - v^*) \tag{5b}$$

Let us further assume that the dependence $\psi^*(\nu)$ can be approximated by a linear function,

$$\psi^*(\nu) - \psi^*(\nu^*) = \left[\frac{d\psi}{d\nu}\right]_{\nu=\nu^*} (\nu - \nu^*).$$ (Here and in the following, we assume that all the derivatives are evaluated at $\nu = \nu^*$ and $\psi = \psi_{eq} = \psi^*(\nu^*)$). Then, taking into account eqs 2b and 2d, we obtain,

$$\frac{dy_V}{dt} = -\frac{y_V}{\tau_{\alpha,eq}} \exp[Q_V y_V] \qquad (6a)$$

$$\frac{dy_H}{dt} = -\frac{y_H}{\tau_{\alpha,eq}} \exp[Q_H y_H] \qquad (6b)$$

Here,

$$Q_V = -(V_I - V^*)\frac{E_2 - E_1}{RT}\left(\frac{d\psi}{d\nu}\right)\left[\frac{\partial V}{\partial \psi}\frac{d\psi}{d\nu} + \frac{\partial V}{\partial \nu}\right]^{-1} \qquad (7a)$$

$$Q_H = -(H_I - H^*)\frac{E_2 - E_1}{RT}\left(\frac{d\psi}{d\nu}\right)\left[\frac{\partial H}{\partial \psi}\frac{d\psi}{d\nu} + \frac{\partial H}{\partial \nu}\right]^{-1} \qquad (7b)$$

Equations 6a and 6b are identical to those proposed by Trachenko and Zaccone[47] based on the concept of local event dynamics; here, they emerge naturally from the TS2 interplay between the fast and slow variables. The expressions in the square brackets in the rhs of eqs 7a and 7b are both negative – as the occupancy $\nu$ is increased, the free volume, specific volume, and enthalpy all decrease. Therefore, for the **_down-jump experiments_** ($T_I > T_a$), $Q_V$ and $Q_H$ are both positive, and the relaxation is stretched-exponential.[35,36,47] For the **_up-jump experiments_** ($T_I > T_a$), $Q_V$ and $Q_H$ are both negative, and the relaxation is compressed-exponential.[36,47] In the limit $T_I \to T_a$, both $Q_V$ and $Q_H$ would approach zero, and the relaxation would become similar to the classical Debye exponential (although, of course, the magnitude of the un-normalized volume or enthalpy deviations would be infinitesimally small).

### 2.3. Stress Relaxation Experiments

Next, let us consider a simple shear stress relaxation experiment. Suppose the material was subjected to a pure shear, with von Mises stress $\sigma_{vM}$, defined as $\sigma_{vM} = \left( \frac{3}{2} \left[ \sigma_{ij} + P\delta_{ij} \right]^2 \right)^{1/2}$, where $\sigma_{ij}$ represents the stress tensor, $P$ is the hydrostatic pressure, and $\delta_{ij}$ is the Kronecker delta. For the sake of simplicity, we assume that $P = 0$, although the approach is easy to generalize for nonzero pressure as well. The free energy $G$ (eq 1) can be then modified as follows (see, e.g., Long et al.[60,61]),

$$A = G + V \left[ \frac{\sigma_{vM}^2}{6\mu} - \sigma_{vM}\gamma \right] \tag{8}$$

Here, $V$ is the volume of a cooperatively rearranging region (CRR), $\mu = \mu(\psi, v)$ is the shear modulus and $\gamma$ is the shear deformation, defined as $\gamma = \sqrt{\frac{2}{3}} \left( \left[ \varepsilon_{ij} + \frac{v_P}{1-2v_P} \varepsilon_{kk}\delta_{ij} \right]^2 \right)^{1/2}$, with $\varepsilon_{ij}$ being the strain tensor, and $v_P$ the Poisson's ratio. Minimizing eq 8 with respect to $\sigma_{vM}$, we obtain the standard linear elasticity relation,

$$\sigma_{vM} = 3\mu\gamma \tag{9}$$

Now, suppose that the external stress is removed but the deformation is maintained constant. In that case, the appropriate form for the free energy is,

$$A = G + V \frac{3}{2} \mu\gamma^2 \tag{10}$$

The stress, measured as a function of time, is then given by,

$$\sigma(t) = 3\gamma\mu(t) \tag{11}$$

To model the function $\mu(t)$, we have to re-define our initial variables. We will now define $\psi^*(v)$ as the solution of equation $\dfrac{\delta A(\psi, v)}{\delta \psi} = 0$, and $v^*$ as the solution of equation $\dfrac{\delta A(\psi^*(v), v)}{\delta v} = 0$. This requires additional assumptions about the functional dependence $\mu(\psi, v)$ -- for now, we will leave it open and only assume that it is a known smooth function.

As in the previous analysis, let us define,

$$y_\sigma = \frac{\sigma - \sigma^*}{\sigma_I - \sigma^*} = \frac{\mu - \mu^*}{\mu_I - \mu^*} \tag{12}$$

We further assume that the material is sufficiently close to its glass transition, so that the modulus relaxation can be measured over realistic experimental timescales. In eq 12, $\sigma = 3\gamma\mu$, $\sigma_I = 3\gamma\mu_I$, $\sigma^* = 3\gamma\mu^*$, $\mu_I = \mu(\psi^*(v_I), v_I)$, and $\mu^* = \mu(\psi^*(v^*), v^*)$. The value of $v_I$ depends on both $\sigma_{vM}$ and the rate of the external loading.

The relaxation dynamics of $y_\sigma$ is then given by,

$$\frac{dy_\sigma}{dt} = -\frac{y_\sigma}{\tau_{\alpha,eq}} \exp[Q_\sigma y_\sigma] \tag{13}$$

$$Q_\sigma = -(\mu_I - \mu^*)\frac{E_2 - E_1}{RT}\left(\frac{d\psi}{dv}\right)\left[\frac{\partial \mu}{\partial \psi}\frac{d\psi}{dv} + \frac{\partial \mu}{\partial v}\right]^{-1} \tag{14}$$

The modulus dependence on $v$ can be elucidated based, e.g., on the analysis of Zaccone and Terentjev[62–64]. If the occupancy $v$ is increased, the average number and "quality" of contacts between the

"constitutive particles" (e.g., monomers) increases, and the modulus is expected to increase as well. Therefore, the expression in the square brackets in the rhs of eq 14 must be positive. What follows is that the trend for the stress relaxation is opposite that of volume or enthalpy relaxation – if $\mu_I - \mu^* > 0$, the relaxation is compressed, while if $\mu_I - \mu^* < 0$, it is stretched. Note, however, that in a "thermomechanical" measurement, where the stress is caused by the temperature change (this type of experiment is important, e.g., for shape memory materials), for the **_down-jump experiments_** (T_I_ > T_a_), $\mu_I - \mu^* < 0$ (higher-temperature modulus is smaller than the lower-temperature one), and the relaxation is stretched-exponential, similar to that of enthalpy and volume; likewise, for the **_up-jump experiments_** (T_I_ > T_a_), $\mu_I - \mu^* > 0$, and the relaxation is compressed-exponential.

One interesting special case here corresponds to $\mu^* = 0$, i.e., to "perfectly plastic" materials (shear storage modulus approaches zero in the limit of zero frequency). In that case, eq 13 can be re-written as,

$$\frac{d\sigma}{dt} = -\frac{\sigma}{\tau_{\alpha,eq}} \exp\left[Q_\sigma \frac{\sigma}{\sigma_I}\right] \tag{15a}$$

or,

$$\frac{d\sigma}{dt} + \frac{\sigma}{\tau_{\alpha,eq}} \exp\left[-\frac{\frac{E_2 - E_1}{RT}\left(\frac{d\psi}{dv}\right)\left[\frac{\partial \mu}{\partial \psi}\frac{d\psi}{dv} + \frac{\partial \mu}{\partial v}\right]^{-1}}{3\gamma}\sigma\right] = 0$$

(15b)

This, in turn, can be re-written as,

$$\frac{d\sigma}{dt} + \frac{\sigma}{\tau_{\alpha,eq}} \exp\left[-\frac{\sigma v^*}{RT}\right] = 0 \tag{16}$$

where we defined

$$v^* = \frac{(E_2 - E_1)\left(\dfrac{d\psi}{d\nu}\right)}{3\gamma\left[\dfrac{\partial\mu}{\partial\psi}\dfrac{d\psi}{d\nu} + \dfrac{\partial\mu}{\partial\psi}\right]} \tag{17}$$

Equation 16 is essentially the well-known Eyring[65] model for plasticity, and the "Eyring activation volume" is then given by eq 17. In our model, the activation volume is not a "material parameter" but depends also on the initial deformation – it would be interesting to see if this prediction is consistent with experiments. The solution of eq 16 at early times is often approximated by the Guiu-Pratt[42] equation,

$$\sigma = \sigma_I - A\ln\left(1 + \frac{t}{B}\right) \tag{18}$$

where $A$ is a constant with units of stress, and $B$ is a constant with units of time (see, e.g., Lazzeri et al.)[43] Below, we will show that such a behavior is indeed consistent with the exact numerical solution of eq 16.

The above discussion was devoted to the shear stress relaxation. However, the same arguments could be made for the tensile or compressive stress relaxation, provided that the initial deformation was elastic (no plastic deformation) and the Poisson's ratio could be treated as a constant over the course of the experiment. In that case, eqs 13, 15a-b, 16, and 17 would still be applicable, although the expression for $Q_\sigma$ (eq 14) will have to be modified.

### 2.4. General Properties of the Relaxation Function and Comparison with the Kohlrausch-Williams-Watts (KWW) Expression

The above discussion showed that various relaxation laws can be modeled using a single equation,

$$\frac{dy}{dt} = -\frac{y}{\tau^*}\exp(Ky) \tag{19a}$$

$$y(0)=1 \qquad (19b)$$

We will refer to this as the Trachenko-Zaccone[47] (TZ) equation, and label its solution as $Y_{TZ}\left(\dfrac{t}{\tau^*};K\right)$. Note that $K$ can be both positive and negative. Note also that the solution of eqs 19a-b can be written in an implicit form using the exponential integral function $Ein(z)=\int_0^z [1-e^{-t}]\dfrac{dt}{t}$, as follows, $t=\tau^*\left[\ln(Ky)-Ein(Ky)+Ein(1)\right]$. In Figure 1, we plot the function $Y_{TZ}(x;K)$ for several values of $K$.

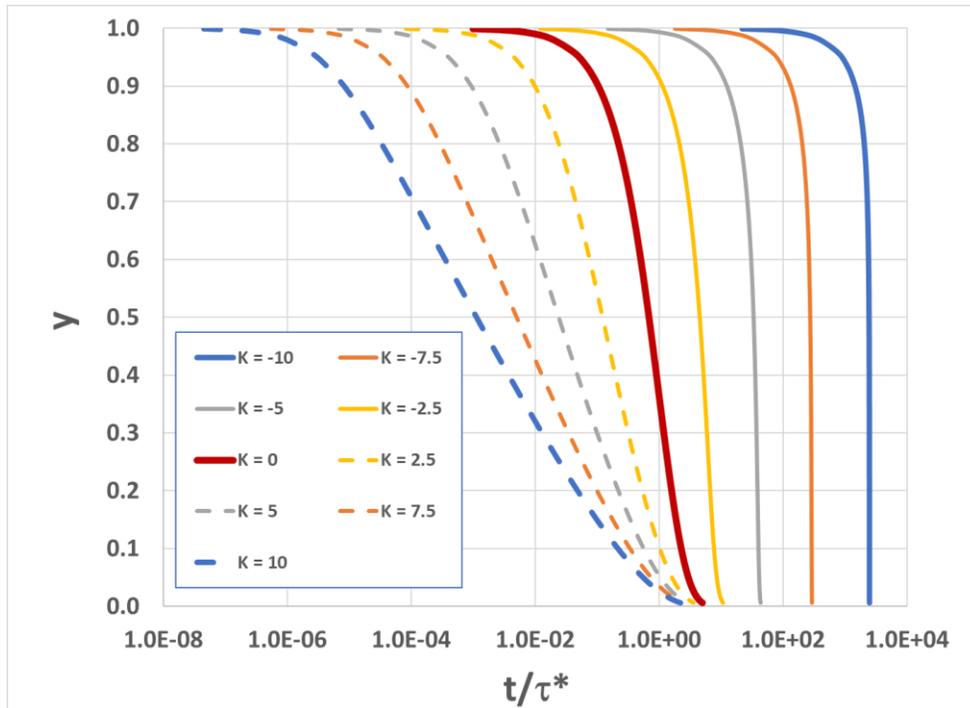

Figure 1. Calculated $Y_{TZ}(t/\tau^*;K)$ for various values of K. The solid lines correspond to compressed exponential behavior (up-jumps, K < 0), while the dashed line depict stretched exponential behavior (down-jumps, K > 0). The thick red line is the Debye relaxation (K = 0).

As already pointed out by Trachenko and Zaccone, the function $Y_{TZ}\left(\dfrac{t}{\tau^*};K\right)$ can be reasonably well approximated by the Kohlrausch-Williams-Watts[37,38] (KWW) function,

$$y = \exp\left[-\left(\dfrac{t}{\tau_{KWW}}\right)^{\beta}\right] \qquad (20)$$

In Figure 2a-b, we plot the dependence of $\beta$ and $\tau_{KWW}/\tau^*$ on $K$.

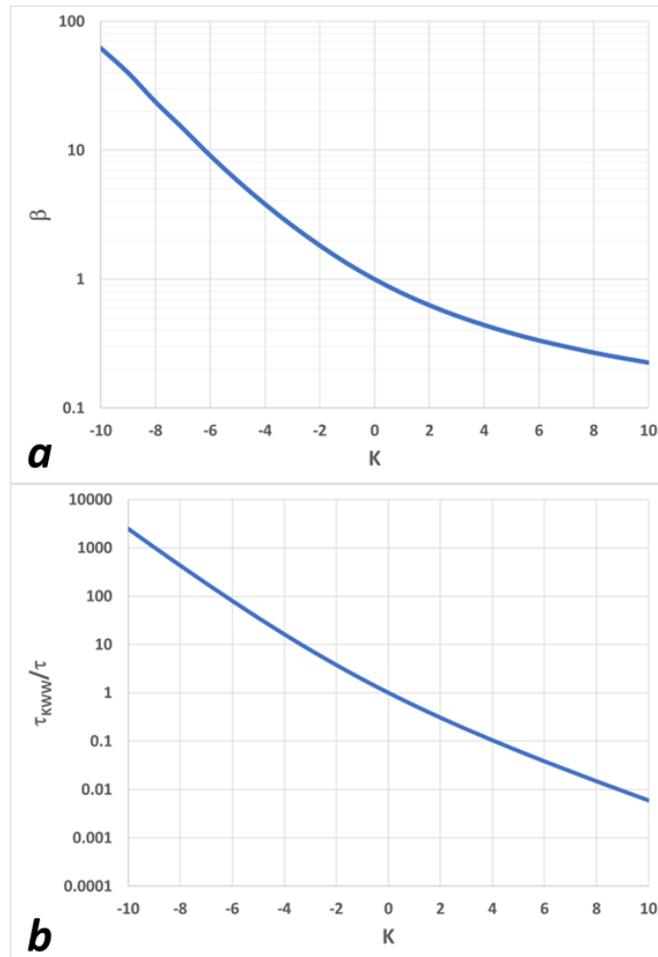

Figure 2. (a) KWW power exponent $\beta$ as function of the TZ parameter $K$; (b) ratio of the KWW relaxation time to the TZ relaxation time as function of the TZ parameter $K$.

Thus, if a relaxation process has been successfully modeled with KWW function, Figure 2 would serve as a "converter" to the TZ function. The relationship is fully reversible. We will use this relationship below when discussing the stress relaxation study.

## 3. Results and Discussion

### 3.1. Comparison with Experiments

#### 3.1.1. Volume Relaxation

As a first example, we consider the volume relaxation experiments of Struik[21], considered to be a "classical" illustration of the "asymmetry of approach". In this study, polystyrene (PS) was quenched from the initial temperature of 110 °C to the final temperature of 89 °C, and the specific volume was then measured as a function of time ("down-jump"); the second experiment involved the rapid heating of the same PS from the initial temperature of 83 °C to the final temperature of 89 °C, again with the measurement of the specific volume as a function of time ("up-jump"). These experiments are typically modeled using TNM[31–33] or KAHR[26] theories, with varying degrees of success (see, e.g., Simon et al. [22,24,28,66]). Here, we illustrate that our TS2 approach successfully describes the Struik experiment, at least semi-quantitatively.

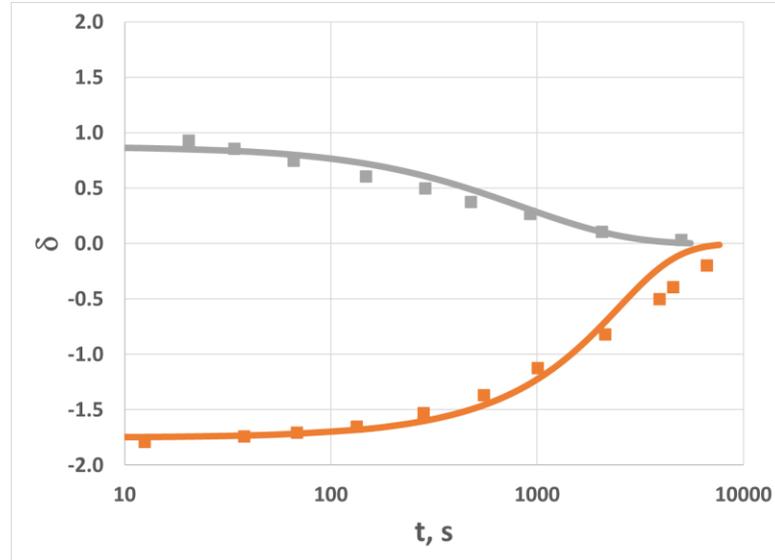

Figure 3. PS volume relaxation: down-jump from 110 °C to 89 °C (grey) and up-jump from 83 °C to 89 °C (grey). The squares are the data of Struik[21] and the lines are the fits using the current model. The model parameters are given in Table 1. See text for more details.

In Figure 3, the normalized difference between the current and the equilibrium specific volume, $\delta = \dfrac{V(t)-V^*}{V^*} \times 10^3$, is plotted as a function of time. We can describe the down-jump with the stretched-exponential KWW function, and the up-jump with the compressed-exponential one. Thus, it is almost trivial that they can also be described by the TZ-functions. To test our theory further, we need to introduce two additional constraints:

1. The relaxation time ($\tau_{\alpha,\,eq}$ in eq 6a) is the same for both sets, given that it represents the equilibrium alpha-relaxation at T = 89 °C.

2. The ratio of the initial magnitude of $\delta$ (labeled $\delta_I$) to the nonlinearity parameter ($Q_V$ in eq 6a) is the same for both sets (see eq 7a).

The lines in Figure 3 correspond to the best-fit results satisfying those constraints. Basically, the fit uses four independent parameters – the relaxation time, $\tau_{\alpha,\,eq}$, for the up-jump, the nonlinearity parameter, $Q_V$, for both the up-jump and the down-jump, and the initial magnitude, $\delta_I$, for the up-jump. The other two parameters (the relaxation time for the down-jump and the initial magnitude for the down-jump) are fixed by the constraints. The parameter values are summarized in Table 1.

*Table 1. Model parameters for the volume relaxation fitting*

| **Experiment** | $\tau_{\alpha,eq}$, s | $\delta_I$ | $Q_V$ |
|---|---|---|---|
| Down-jump | 1200 | 0.875 | 0.5 |
| Up-jump | 1200 | -1.75 | -1 |

The relaxation time of ~1200 s is reasonable, given that the aging temperature (89 °C) is several degrees below the glass transition temperature (for PS, it is about 95 – 100 °C, depending on the molecular weight and tacticity[67,68]). The initial specific volume should, in principle, correlate with the temperature difference between the initial and final temperature, but there are two complications. For the down-jumps, the initial relaxation of the specific volume occurs simultaneously with the cooling process, so by the time the true isothermal aging starts, a portion of the volume relaxation has already taken place. For the up-jumps, the initial state of the material may or may not be fully equilibrium, in which case the "fictive temperature" can be higher than the actual one. Thus, in this analysis, we treat the initial specific volume values as free parameters, although in principle, they could be estimated at least with reasonable accuracy. Finally, the nonlinearity parameters are treated as free parameters here; in the future, they would be related back to the SL-TS2 thermodynamic coefficients.

### 3.1.2. Enthalpy Relaxation

The second example is the enthalpy relaxation, also in PS.[28] The enthalpy relaxation following up- and down-jumps is measured using DSC, and converted to the relaxation of the fictive temperature, based on the following definition,

$$H(t) - H^* = C_P\left(T_f(t) - T_a\right) \qquad (21)$$

Figure 4 illustrates two up-jumps (by 10 °C and by 4 °C) and two down-jumps (by the same temperature increments) to the same aging temperature, $T_a$ = 107 °C. In this study, the final temperature is close to – or even above – the glass transition. Thus, the relaxation time is expected to be shorter than in the previous example, but other than that, the analysis fundamentally remains the same. The model parameters are summarized in Table 2. Our results are comparable with those of Grassia et al.[28] who developed a modified TNM model and successfully reproduced both asymmetry of approach and memory experiments. We intend to apply our model to other protocols (uniform cooling; memory experiment etc.) in the future.

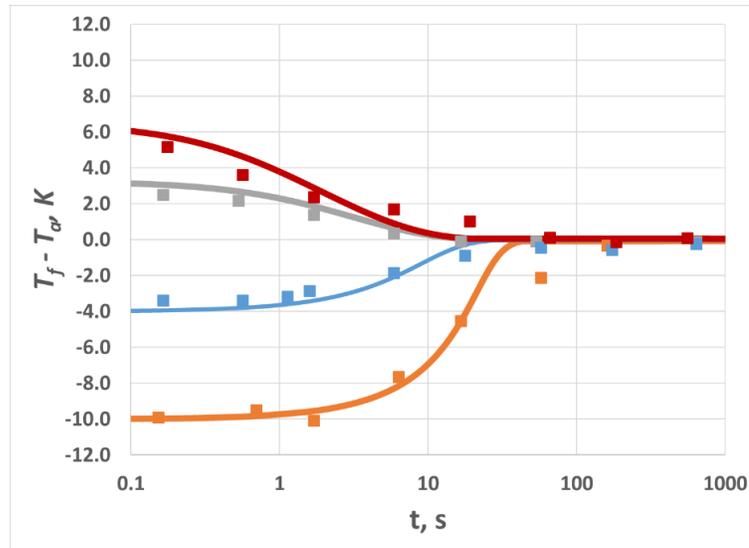

*Figure 4. Enthalpy relaxation of PS – down-jumps of 10 °C (red) and 4 °C (grey); up-jumps of 10 °C (orange) and 4 °C (blue). In all cases, the aging temperature is 107 °C. The squares are data from Grassia et al.[28], and the lines are the fits using the current model. The model parameters are given in Table 2. See text for more details.*

*Table 2. Model parameters for the enthalpy relaxation fitting.*

| Experiment | $\tau_{\alpha,eq}$, s | $(\Delta T)_l$, K | $Q_v$ |
|---|---|---|---|
| Down-jump 10 °C | 5 | 6.5 | 1.3 |
| Down-jump 4 °C | 5 | 3.25 | 0.65 |
| Up-jump 4 °C | 5 | -4 | -0.8 |
| Up-jump 10 °C | 5 | -10 | -2 |

### 3.1.3. Stress Relaxation

In the final example, we consider the stress relaxation experiments of Liu et al.[69] In that study, the authors measured the stress vs. time for polystyrene PS-45K ($M_w$ = 45 kg/mol) at several temperatures (T = 45, 50, and 55 °C). They showed that the compression stress relaxation curves could be well described by the KWW relaxation function with $\beta \approx 0.5$ (see Table 3 for details).

As discussed above, the KWW function can be interpreted as an approximate solution of the nonlinear TZ equation 15a. Furthermore, eq 15a can be re-interpreted as a combination of an Eyring dashpot and elastic spring, so the early-time asymptotic behavior would be given by the Guiu-Pratt logarithmic expression,

$$\sigma = \sigma_I - \frac{RT}{v^*}\ln\left(1 + \frac{t}{c}\right) \qquad (22)$$

Here, $v^*$ is activation volume, and $c$ is a characteristic time.

Going back to eq 16 above, we can re-write it in the form of the "standard" TZ equation (eq 19) after defining $y = \sigma/\sigma_I$. In that case, eq 22 is the early-time asymptotic of the solution of the TZ equation; the general solution (including both early and late times) can be written as,

$$\sigma(t) = \sigma_I Y_{TZ}\left(\frac{t}{\tau^*}; K\right), \qquad (23)$$

with $K = \frac{\sigma_0 v^*}{RT} - B$, $B$ = 1.24, and $\tau_{\alpha,eq} = (K+B)e^K c = \frac{\sigma_0 v^*}{RT}e^{\left(\frac{\sigma_0 v^*}{RT}-B\right)}c$.

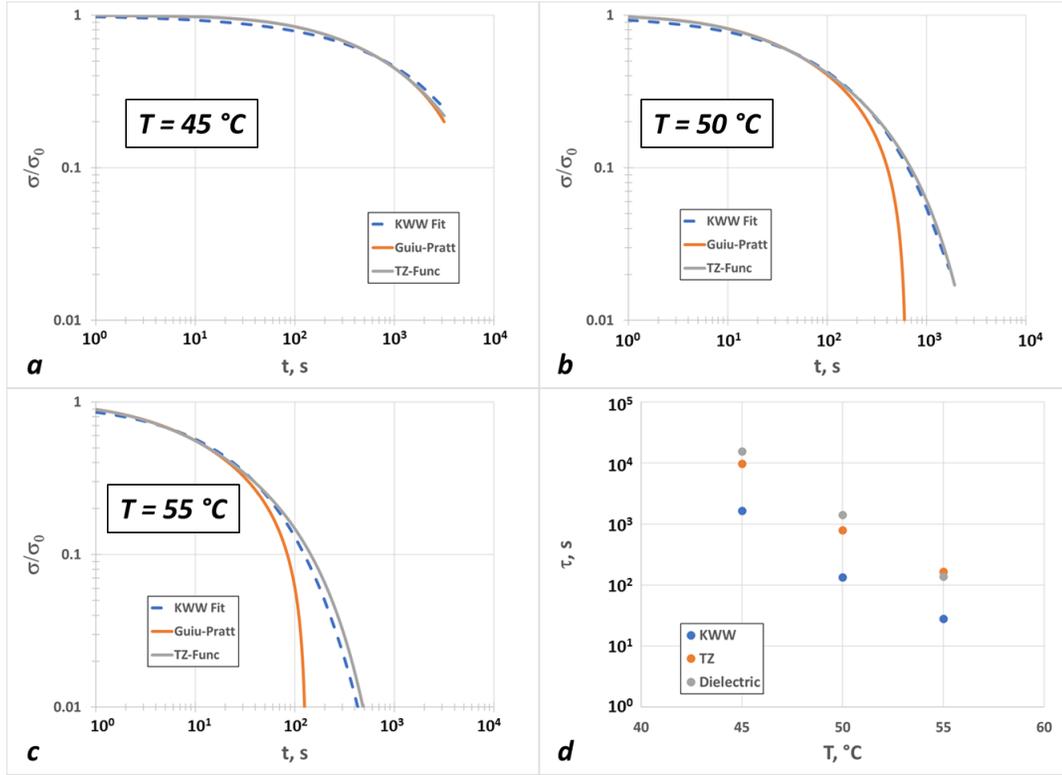

*Figure 5. Normalized stress relaxation curves of PS-45K at (a) T = 45 °C; (b) T = 50 °C; (c) T = 55 °C. For all three temperatures, blue dashed lines are the KWW fit to data from Ref. [69]; grey lines are the best-fit numerical TZ solutions, and orange lines are the corresponding Guiu-Pratt early-time asymptotic solutions. (d) The KWW relaxation times (Ref. [69]), the measured dielectric α-relaxation times (Ref. [69]), and the estimated equilibrium α-relaxation time based on TZ-fit. The model parameters are given in Table 3. See text for more detail.*

In Figure 5, we plot the results of the stress relaxation modeling for PS-45K at various temperatures. For T = 45 °C (Figure 5a), T = 50 °C (Figure 5b), and T = 55 °C (Figure 5a), the original authors' KWW fits to the data[69] are plotted as the blue dashed lines; the exact TZ fits are shown as grey lines, and the approximate early-stage logarithmic fits are shown as orange lines. For all the temperatures, TZ approximates KWW very well, while the logarithmic fit works at early times where $\sigma/\sigma_I > 0.3$. To our knowledge, this duality in the description of stress relaxation in glassy polymers has not yet been discussed in the literature and is worth further analysis. The model parameters for this study are summarized in

Table 3. The column "Time Ratio" is the calculated ratio of the KWW time to the TZ time, based on the value of $Q_\sigma$ (see Figure 2b, with K = $Q_\sigma$). We can thus estimate the TZ-relaxation time which is, within our model, equal to the equilibrium α-relaxation time. The KWW and TZ relaxation times are plotted as functions of temperature (Figure 5d), together with the dielectric relaxation data also provided by Liu et al.[69] It can be clearly seen that the TZ characteristic relaxation times are much closer to the dielectric data than the KWW ones. We also estimated the apparent Eyring activation volume, $v^*$, that one would calculate by assuming that the logarithmic decay of the stress is due to the Eyring-Guiu-Pratt mechanism. The calculated values are reasonably close (within a factor of 3) to literature estimates.[70]

*Table 3. Parameters for the PS-45K stress relaxation study. The parameters in gold are from Ref. [69] while those in green are the result of the current analysis. For more detail, see text.*

| T, °C | β | $\tau_{KWW}$, s | $\tau_{diel}$, s | $\sigma_l$, MPa | Time Ratio | $\tau_{\alpha,eq}$, s | $Q_\sigma$ | $v^*$, nm$^3$ |
|---|---|---|---|---|---|---|---|---|
| 45 | 0.51 | 1640 | 15594 | 28 | 0.169 | 9731 | 3.1 | 0.49 |
| 50 | 0.53 | 134 | 1411 | 25 | 0.169 | 795 | 3.1 | 0.55 |
| 55 | 0.56 | 28 | 137 | 22 | 0.169 | 166 | 3.1 | 0.64 |

### 3.2. Discussion

In the above analysis, we proposed a unified approach to the description of various types of relaxation (volume, enthalpy, stress, and possibly dielectric) in amorphous materials near the glass transition temperature. This approach is based on three assumptions. First, it is assumed that the Gibbs (or another appropriate) free energy of the material can be written as a function of two state variables, $\psi$ and $v$, where $\psi$ is related to the "internal state" of the material (fraction of the "solid" or "rigid" elements), and $v$ is related to the fractional free volume. Second, it is assumed that the characteristic relaxation time for $\psi$ is the beta-relaxation time ("fast"), and the characteristic relaxation time for $v$ is the alpha-relaxation time ("slow") which is itself a function of $\psi$. Finally, it is assumed that the variations of $v$ and $\psi$ between the initial and final state are small enough to stipulate a simple proportionality between them and thus to adopt that the derivative $\frac{d\psi}{dv}$ as a constant. These assumptions allow us to formulate a unified framework for the description of the volume and enthalpy relaxation, as well as – with some minor modification – stress relaxation.

Mathematically, the proposed approach leads to the re-formulation of the dynamics as a single nonlinear first-order differential equation (Trachenko-Zaccone[47] equation) with respect to $v$ for every particular relaxation process. Thus, it can be treated as a modified free-volume theory[71–76] – or as a modified "material time" theory (see, e.g., Niss et al. [36] who arrived at a similar equation using different arguments). The solutions of the TZ equation depend on the sign of the nonlinearity parameter *K* – for any positive *K*, the relaxation is stretched-exponential, for *K* = 0, it is exponential, and for any negative *K*, it is compressed-exponential. For the case of stress relaxation, the above analysis leads naturally to the Eyring-like model with the Guiu-Pratt logarithmic decay of the stress as a function of time. Other functional forms, such as the KWW function, have been also used to parameterize the stress relaxation of various polymers.[77] As we discussed, logarithmic Guiu-Pratt, KWW stretched-exponential, KWW compressed-exponential, and Debye functions, are all recovered as special limits of the TZ solutions.

The proposed framework is, of course, a simplified mean-field, phenomenological analysis based on a statistical thermodynamics model. In reality, each macroscopic system is characterized by the distribution of relaxation times, microscopic states, and local energies. It would be instructive to see what the dependence of the relaxation time on aging time tells us about the long-range interactions, fractal structures, etc.

We expect that the proposed model can be adapted further to describe other types of mechanical tests (e.g., constant strain-rate tensile or shear stress-strain curves, measurements of fatigue under repeated tensile or shear cycles, shape memory tests, etc. – see, e.g., review by Medvedev and Caruthers[78] and references therein). To do this, we would need to stipulate the dependence of the shear and bulk moduli on the state variables, $\psi$ and $\nu$. This is another topic for future work.

## 4. Conclusions

We developed a framework to describe relaxation phenomena (of stress, specific volume, and enthalpy) in glassy materials. Within this framework, the evolution of the "slow" state variable, $\nu$, is described by the nonlinear Trachenko-Zaccone equation whose solutions include logarithmic, KWW (both stretched and compressed), and simple exponential functions. The variable $\nu$ itself is related to the fractional free volume, and the response variables can be expressed as linear functions of $\nu$ if the magnitudes of their changes are small enough.

The new framework is applied to several examples from literature – one for specific volume, one for enthalpy, and one for shear stress relaxation. The agreement between the unified theory and experiments is good and comparable to that of the individual earlier models. The parameters used in modeling relaxations will be used in calibrating the free energies and EoS for various glassy polymers. Once those calibrations are done, the theory will be tested for more complex thermal and mechanical histories. All in all, the presented unifying framework provides the unprecedented possibility of rationalizing widely different relaxation behaviors in glasses based on a unifying physical model.

## Acknowledgment

A.Z. gratefully acknowledges funding from the European Union through Horizon Europe ERC Grant number: 101043968 ``Multimech'', and from US Army Research Office through contract nr. W911NF-22-2-0256. V.G. thanks the Faculty of Mechanical Engineering, Technion, Israel, for hospitality and support during his sabbatical. We would like to thank Luigi Grassia (University of Campania) and Grigori Medvedev (Purdue University) for stimulating discussions.

# Graphical Abstract

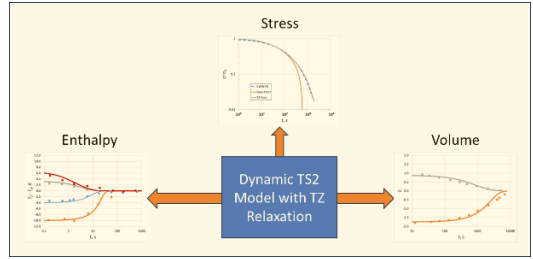